\documentclass[aps,prl,twocolumn,superscriptaddress]{revtex4}
\usepackage{amssymb}
\usepackage{graphicx}
\usepackage{amsmath}
\usepackage{verbatim}

\begin{document}
	
	\title{Energy absorption spectroscopy of unitary Fermi gases in a uniform potential}
	\author{Pengfei Zhang}
	\affiliation{Institute for Advanced Study, Tsinghua University, Beijing 100084, China}
	
	\author{Zhenhua Yu}
	\email[]{huazhenyu2000@gmail.com}
	\affiliation{Laboratory of Quantum Engineering and Quantum Metrology, School of Physics and Astronomy, Sun Yat-Sen University (Zhuhai Campus), Zhuhai 519082, China}
	\affiliation{State Key Laboratory of Optoelectronic Materials and Technologies,
Sun Yat-Sen University (Guangzhou Campus), Guangzhou 510275, China}
	\date{\today }
	
	\begin{abstract}
	We propose to use the energy absorption spectroscopy to measure the kinetic coefficients of unitary Fermi gases in a uniform potential. We show that in our scheme, the energy absorption spectrum is proportional to the dynamic structure factor of the system. The profile of the spectrum depends on the shear viscosity $\eta$, the thermal conductivity $\kappa$ and the superfluid bulk viscosity $\xi_3$. We show that extraction of these coefficients from the spectrum is achievable in present experiments.
	\end{abstract}
	
	\maketitle
Strong interaction remains a main challenge to modern many-body theory. Recent applications of 
holographic duality provide a new way to explore physics of strongly interacting systems \cite{Son Review, Sachdev}.
The principle of holographic duality is to transcribe a strongly interacting quantum system in $d+1$ space-time dimensions into a classical gravity theory in $(d+1)+1$ dimensions. 
In particular, the method has been used to calculate hydrodynamic behaviors in the long wave-length and low energy limit \cite{Son Review,Sachdev}. 

One example is that for a wide class of interacting systems, the diffusion coefficient $D_\eta\equiv\eta/s T$, as $\eta$ is the shear viscosity and $s$ is the entropy density, has been shown to have a universal value $1/4\pi T$ (the light speed $c=1$) \cite{DTSon}. Since the ratio $\eta/s$ is known to diverge in the noninteracting limit, Kovtun, Son and Starinets conjectured the existence of a lower bound for the ratio, i.e., $\eta/s\geq 1/4\pi$ \cite{DTSon}.
Lately this lower bound has been generalized to the form $D\gtrsim v^2/T$, where $D$ is the diffusion coefficient for any diffusive mode and $v$ is a typical velocity scale of the system \cite{lower Hartnoll,lower Blake}. In addition, an upper bound has also been put forward via considering the constrains from causality: $D\lesssim \tau_{\text{eq}}v_{\text{LR}}^2$ \cite{upper Hartnoll,upper Lucas}. In this case, one assumes the Lieb-Robenson velocity $v_{\text{LR}}$ as the upper bound for propagation speeds of any local perturbations, and $\tau_{\text{eq}}$ is the thermalization time to reach local equilibrium.

The unprecedented power of tuning interactions in ultracold atomic gases makes the system an ideal platform to test the above predictions. 
Previous measurements of the shear viscosity for unitary Fermi gases find a minimum value of the ratio $\eta/s$ about several times the lower bound $1/4\pi$ \cite{Thomas2011sci,Thomas2011njp,Thomas2014}.
However, accurate extraction of the kinetic coefficients in the experiments are hampered by the inhomogeneity of the gases confined in external harmonic traps. Recent realization of uniform potentials for Bose 
\cite{Hadzibabic2013,Hadzibabic2015,Hadzibabic2016} and Fermi gases \cite{Zwierlein2017} offers a way to circumvent this difficulty.

In this work, we propose an energy absorption spectroscopic method to measure the kinetic coefficients of unitary Fermi gases in a uniform potential. 
During the early days to produce ultracold atomic gases, the noise in the harmonic trap used to confine the gases in experiment gives rise to heating and was considered an obstacle for achieving low temperatures. Later, experimentalists observed resonant parametric heating in 
weakly interacting gases when the deliberately introduced modulation of frequency $\omega$ to the harmonic potential of trapping frequency $\omega_{\rm ho}$ is tuned at $\omega=2\omega_{\rm ho}$ \cite{thomas, thomas98}.
With a long wave-length and low frequency perturbation added on top of the uniform potential, we show how the energy absorption spectrum, which is related to the dynamic structure factor, is determined by the hydrodynamic equations; consequently the kinetic coefficients, such as the shear viscosity $\eta$, the thermal conductivity $\kappa$ and the superfluid phase bulk viscosity $\xi_3$, can be extracted from the features of the spectrum. With the estimate of the magnitudes of the kinetic coefficients by the linearized Boltzmann kinetic equation \cite{Landau kinetics,schafer kappa}, we show that the effects of the kinetic coefficients on the spectrum can be observed within current experimental resolution.
Our proposed measurement gives an opportunity to determine the kinetic coefficients more accurately than those previously from inhomogeneous gases, and would put the theoretical predictions regarding strong interacting systems to a more stringent test.

We consider that on top of the uniform potential, there is a spatial and temporal potential perturbation $\delta U(\mathbf r,t)=2 U_0\cos(\mathbf q\cdot\mathbf r-\omega t)$ in the long wavelength and low frequency limit such that $q\ll k_F$ and $\omega\ll E_F$. Here the Fermi momentum is $k_F\equiv (6\pi^2N/V)^{1/3}$ and the Fermi energy is $E_F\equiv k_F^2/2m$, with $N$ the number of fermions of each component of the unitary Fermi gas and $V$ the system volume. Such a perturbation, for example, can be conveniently realized by a digital micromirror device (DMD) \cite{MacAulay1998, Jovin1999}. The resultant perturbation Hamiltonian is $\hat H_{\rm per}=\int d\mathbf r \delta U(\mathbf r,t)\hat n(\mathbf r)$, with $\hat n(\mathbf r)$ the number density operator. The unitary Fermi gas would absorb energy from the time varying perturbation; at the level of linear response, the energy absorption spectrum is proportional to
	\begin{align}
	\Gamma(\mathbf q,\omega)=&2\pi\omega(1-e^{-\omega/T}) \nonumber\\
	&\times\sum_{f,i}P_i |\langle f|\hat n_{\mathbf q}|i\rangle|^2\delta(\omega+E_i-E_f),\label{lr}
	\end{align}
where $\hat n_{\mathbf q}\equiv\int d\mathbf r \hat n(\mathbf r) e^{-i\mathbf q\cdot\mathbf r}$, 
$|i\rangle$ are the eigenstates of the system, $E_i$ are the corresponding eigenenergies, and $P_i$ is the thermal distribution. 
Exactly speaking, the uniform potentials realized in the experiments come with walls at boundaries and momentum is not conserved. For simplicity, we consider in the thermodynamic limit that boundary effects shall be negligible. 
We take $\hbar=1$, $k_B=1$ throughout. Note that $\Gamma(\mathbf q,\omega)$ is nonnegative and $\Gamma(\mathbf q,\omega)=\Gamma(\mathbf q,-\omega)$.
Thus by our specific perturbation $\delta U(\mathbf r,t)$, the energy absorption spectrum is given by 
\begin{align}
	\Gamma(\mathbf q,\omega)= 2\pi\omega(1-e^{-\omega/T}) S(\mathbf q,\omega)\label{gammas},
	\end{align} 
where $S(\mathbf q,\omega)$ is the the dynamic structure factor. Bragg spectroscopy has been widely used in ultra-cold atomic gases to access the information of the dynamic structure factor \cite{Phillips1999,Ketterle1999,Cornell2008,Vale2009,Sengstock2010,Vale2012,Vale2016}. A recent experiment designed to probe 	
the Goldstone mode and pair breaking excitations in the center part of harmonically trapped unitary Fermi gases has managed to extend the transferred momentum to as low as $q\sim 0.5 k_F$ \cite{Vale2017}. Quasiparticle energies and quantum depletion of Bose-Einstein condensates in uniform potentials have been measured at a similar wave-length \cite{Hadzibabic2017a, Hadzibabic2017b}. Our proposed energy absorption spectroscopy implemented by the digital micromirror devices has the prospect to conveniently explore longer wave-length regimes \cite{MacAulay1998, Jovin1999}.

	It is worth mentioning that by coupling different forms of perturbation potentials to the atomic gases, one is able to use the energy absorption spectrum to probe different information of the gases. Previously by the technique of synthetic gauge fields, one is able to generate an effective gauge field $\mathbf A(\mathbf r,t)$ \cite{Spielman2009,Spielman2011}; by varying the effective field $\mathbf A(\mathbf r,t)$, one can use the energy absorption spectrum to measure current-current correlations. For gases in optical lattices, the energy absorption spectrum due to the amplitude and phase modulation of the optical lattice potential has also been shown to give access to the information both of the Mott-insulating gap and to the kinetic-energy correlations \cite{Esslinger2004,Esslinger2004b,Giamarchi2006,Giamarchi2006b,Giamarchi2006c,Esslinger2008,Strohmaier2010,Demler2009,Giamarchi2011}.

For our specific potential perturbation, the energy absorption spectrum (\ref{lr}) of the unitary Fermi gas is related to the density correlation of the system, whose long wave-length and low energy behavior can be derived from the two-fluid hydrodynamic equations \cite{Landau Hydro}. As we are interested in the linear response regime, the linearized hydrodynamic equations, which govern the time evolution of the mass density $\rho$, the mass current density $\mathbf j$, the entropy density $s$ and the superfluid velocity $\mathbf v_s$, are given by \cite{Landau Hydro}
	\begin{align}
	&\frac{\partial \rho}{\partial t}+\nabla\cdot\mathbf j=0,\label{rho}\\
	&\frac{\partial j_a}{\partial t}+\partial_b\Pi_{a b}=-\partial_b\Pi'_{a b}-\rho\partial_aU\label{Newton},\\
	&\frac{\partial s}{\partial t}+\nabla\cdot(s\mathbf v_{n})=\kappa\frac{\nabla^2T}{T},\label{s}\\
	&\frac{\partial \mathbf v_{s}}{\partial t}=-\nabla\mu-\nabla \delta U-\nabla\varphi'\label{vs},
	\end{align}
where the sum of the superfluid density $\rho_s$ and the normal density $\rho_n$ is the total mass density $\rho=mn$, $\mathbf v_n$ is the normal motion velocity, the sum of the superfluid current $\rho_s\mathbf v_{s}$ and the normal current $\rho_n\mathbf v_{n}$ is the total current $\mathbf j$, and the non-dissipative part of the energy-momentum tensor density is $\Pi_{ab}=p\delta_{ab}$ with $p$ the pressure. The relative velocity between the normal and superfluid motion is $\mathbf w=\mathbf v_{n}-\mathbf v_{s}$, and $\mu$ is the chemical potential, and $\delta U$ is the potential perturbation that we introduce to extract the energy absorption rate. The two dissipative terms are $\varphi'=\xi_3\nabla\cdot(\rho_s\mathbf w)+\xi_1\nabla\cdot \mathbf v_{n}$ and $\Pi'_{ab}=-\eta(\partial_bv_{n,a}+\partial_av_{n,b}-\frac{2}{3}\delta_{ab}\nabla\cdot \mathbf v_{n})-\xi_1\delta_{ab}\nabla\cdot(\rho_s\mathbf w)-\xi_2\delta_{ab}\nabla\cdot\mathbf v_{n}$. In addition to the thermal conductivity $\kappa$, the shear viscosity $\eta$ and the bulk viscosity $\xi_2$, the relative motion between the superfluid and the normal part shall give rise to extra dissipation characterised by two viscosities $\xi_1$ and $\xi_3$. The positive definiteness of entropy production requires $\xi_1^2<\xi_2\xi_3$ \cite{Landau Hydro}.
One simplication for unitary Fermi gases is that since at the unitarity point the $s$-wave scattering length $a_s$ is divergent, i.e., 
$1/a_s=0$, the absence of any interaction length scales renders the systems conformal invariant. As a result, $\xi_1=\xi_2=0$ \cite{Son2007}.
Therefore, the dissipation of unitary Fermi gases is captured by the kinetic coefficients $\eta$, $\kappa$ and $\xi_3$, while $\xi_3$ drops out in the normal phase.

	The perturbation $\delta U(\mathbf r,t)$, which enters into Eq.~(\ref{vs}), induces a density fluctuation $\delta\rho(\mathbf r,t)$ over the uniform background. We define 
	the linear response function $\chi(\mathbf{q},\omega)=\delta\rho(\mathbf{q},\omega)/\delta U(\mathbf{q},\omega)$.
	By the fluctuation-dissipation theorem, we find
	\begin{align}
	\Gamma(\mathbf{q},\omega)/V=4\pi\omega e^{-\omega/T}\chi''(\mathbf{q},\omega)/m^2\label{g-chi}
	\end{align}
	with $\chi''(\mathbf{q},\omega)\equiv-\text{Im}[\chi(\mathbf{q},\omega)]$, 
	correspondingly $m^2S(\mathbf{q},\omega)/V=2\chi''(\mathbf{q},\omega)/[\exp(\omega/T)-1]$.

	Since the overall density fluctuation $\delta \rho$ only involves longitudnal modes, we simplify Eqs.~(\ref{Newton}) and (\ref{vs}) by taking the divergence of them, and have in the Fourier transformed form
	\begin{align}
	&\left(\frac\omega q\right)^2\delta \rho=\delta p + \rho \delta U-i\frac{4\eta}{3}\mathbf q\cdot \mathbf v_{n},\label{s1}\\
	&-\left(\frac{\omega}q\right)\hat{\mathbf q}\cdot(\mathbf w-\mathbf v_{n})=\delta\mu+\delta U+i\xi_3\rho_s\mathbf q\cdot\mathbf w.\label{s2}
	\end{align}
Here the symbol $\delta$ denotes fluctuations on the homogeneous background. We have used the continuum equation (\ref{rho}). 
On the other side,  the Fourier transformed forms of Eqs.~(\ref{rho}) and (\ref{s}) become
\begin{align}
&\omega\delta\rho+\rho_n \mathbf q\cdot\mathbf v_n+\rho_s \mathbf q\cdot\mathbf v_s=0,\label{s3}\\
&\omega\delta s+s\mathbf q\cdot\mathbf v_n=\kappa q^2\delta T/T\label{s4}.
\end{align}
Combining Eqs.~(\ref{s1}) and (\ref{s4}) together, one can solve $\chi(\mathbf{q},\omega)$ analytically.
	
	The two-fluid hydrodynamic equations were originally established to explain the superfluid properties of liquid helium-4 \cite{Landau Hydro}. 
It is known that in the normal phase, there is only one propagating longitudinal mode, the first sound, whose velocity $c$ is given by $c^2=\left(\partial p/\partial \rho\right)_{\tilde{s}}$ with $\tilde{s}$ the entropy for unit mass; the heat conduction mode is diffusive. In the superfluid phase, when $\rho_s\neq0$, the heat conduction transforms into the other propagating longitudinal mode, the second sound, whose velocity is $u_2=\tilde{s}\sqrt{\frac{T\rho\rho_s}{cv\rho_n}}$ with $v$ the volume for unit mass. Each of these modes is expected to give rise to a peak in $\chi''(\mathbf q,\omega)$ in the frequency domain, and the dissipations $\sim\eta,\kappa,\xi_3$ determine the peak widths. Indeed, in analysing the data for the dynamic structure factor of liquid helium-4 in the low frequency regime \cite{Helium}, it was found that $S(\mathbf q,\omega)$ can be well approximated by the sum of two parts which each correspond to the two modes; each part takes the form $S_i(\mathbf q,\omega)=(a_i+b_i\omega^2)/[(\omega^2-c_i^2)^2+d_i\omega^2]$. This approximation in the vicinity of the peaks agrees with the analytic expression of the dynamic structure factor derived by Hohenberg and Martin \cite{Hohenberg}. 
 Recently the second sound has been observed in the superfluid phase of unitary Fermi gases as well \cite{Grimm2013}. Here we apply the same approximation to the calculation of $\chi''(\mathbf q,\omega)$ of the unitary Fermi gases and find 
	\begin{align}
	&\chi''_{\rm app}(\mathbf{q},\omega)=\chi''_1(\mathbf{q},\omega)+\chi''_2(\mathbf{q},\omega),\label{chiapp}\\
	&\chi''_1(\mathbf{q},\omega)=\frac{2\omega\rho}{c^2}\frac{\Gamma_1 c^2 q^2}{4\Gamma_1^2 \omega^2+(\omega^2-c^2q^2)^2},\label{chi1}\\
	&\chi''_2(\mathbf{q},\omega)=\frac{2\omega\rho}{c^2}\left(\frac{c_p}{c_v}-1\right)\frac{\Gamma_\eta\Omega^2+\Gamma_\kappa\omega^2}{(\Omega^2-\omega^2)^2+4(\Gamma_\eta+\Gamma_\kappa)^2\omega^2},\label{chi2}
	\end{align}
	where 
	\begin{align}
	&2\Gamma_1=\left[\frac{4\eta}{3\rho}+\frac{\kappa}{c_p}\left(\frac{c_p}{c_v}-1\right)\right]q^2,\label{g1}\\
	&\Omega^2=u_2^2q^2+4\Gamma_\eta\Gamma_\kappa,\\
	&2\Gamma_\kappa=\frac{\kappa q^2}{c_p},\\
	&2\Gamma_\eta=\frac{\rho_s q^2}{\rho_n}\left(\frac{4\eta}{3\rho}+\rho\xi_3\right)\label{ge},
	\end{align}
and $c_v$ and $c_p$ are the heat capacities for unit volume, i.e., $ c_{v(p)}=\rho T(\partial \tilde{s}/\partial T)_{\rho(p)}$. Note that in determining $\chi''_2(\mathbf q,\omega)$, we have expanded to first order of $(c_p-c_v)/c_v$, which is justifiable for incompressible liquids. For unitary Fermi gases in the low temperature regime where phonons are the dominant low energy excitations, $c_v\sim T^3$ and $c_p-c_v=(T/\rho)(\partial p/\partial T)_\rho^2(\partial\rho/\partial p)_T\sim T^7$; this expansion is also justified. In the normal phase where $\rho_s\rightarrow0$, $\chi''_2(\mathbf{q},\omega)/\omega$ becomes the Rayleigh peak of incompressible liquids \cite{principle of condensed matter}. Therefore, if in the experiment two well separated peaks are observed in the energy absorption spectrum, fitting experimental data by Eqs. (\ref{g-chi}) and (\ref{chiapp}) to (\ref{chi2}) could extract the values of the kinetic coefficients $\eta,\kappa,\xi_3$. However, when dissipation is sufficiently strong, the widths of the two peaks may exceed the separation between the peaks. In this situation, the above two-peak approximation breaks down, and one shall solve $\chi''(\mathbf q,\omega)$ via Eqs.~(\ref{s1}) to (\ref{s4}) instead. It is known that $\chi''(\mathbf q,\omega)$ of unitary Fermi gases satisfies the identity $\eta=\lim_{\omega\to0}\lim_{\mathbf q\to0}3\omega^3\chi''(\mathbf q,\omega)/q^4$ \cite{Taylor}.

	The full profile of the energy absorption spectrum $\Gamma(\mathbf q,\omega)$ depends on both the equation of state and the kinetic coefficients.
	Since the interaction in unitary Fermi gases is fine tuned at the unitary limit where $1/a_s=0$, the thermodynamic functions of the unitary Fermi gases are universal \cite{Jason}. Namely, we define $x=\mu/T$ and can express the pressure and the superfluid density as \cite{Yanhua}
	\begin{align}
	p=\frac{T}{\lambda_T^3}f_p(x),\\
	\rho_s=\frac{1}{\lambda_T^3}f_s(x).
	\end{align}
	Here	$\lambda_T$ is the thermal wave-length; the two universal functions $f_p(x)$ and $f_s(x)$ determine the thermodynamic equilibrium properties of the unitary Fermi gas. The exact forms of $f_p(x)$ and $f_s(x)$ can be analytically derived in the low and high temperature limits. The superfluid transition temperature $T_C$ has been found experimentally to be around $0.15T_F$ ($T_F=E_F$) \cite{Zwierlein2012}. In the regime $T/T_C\ll1$, the dominate excitations are phonons. To the lowest order, one could approximate the action of the system by that of non-interacting thermal phonons \cite{effective phonon}. In the high temperature regime, the fermion fugacity $z=\exp(\mu/T)$ is small and one can calculate the thermodynamic functions by the virial expansion \cite{Jason, Ho2004,Liu2009,Yu2009,Liu2013,
chao castin}. Experimentally, previous measurement of the equation of state yields the general behavior of $f_p(x)$ \cite{Zwierlein2012}, and the superfluid density $\rho_s$ in a wide temperature range below the critical temperature has also been extracted out from the previous second sound experiment \cite{Grimm2013}. 
				
		\begin{figure}[t]
			\includegraphics[width=3 in]{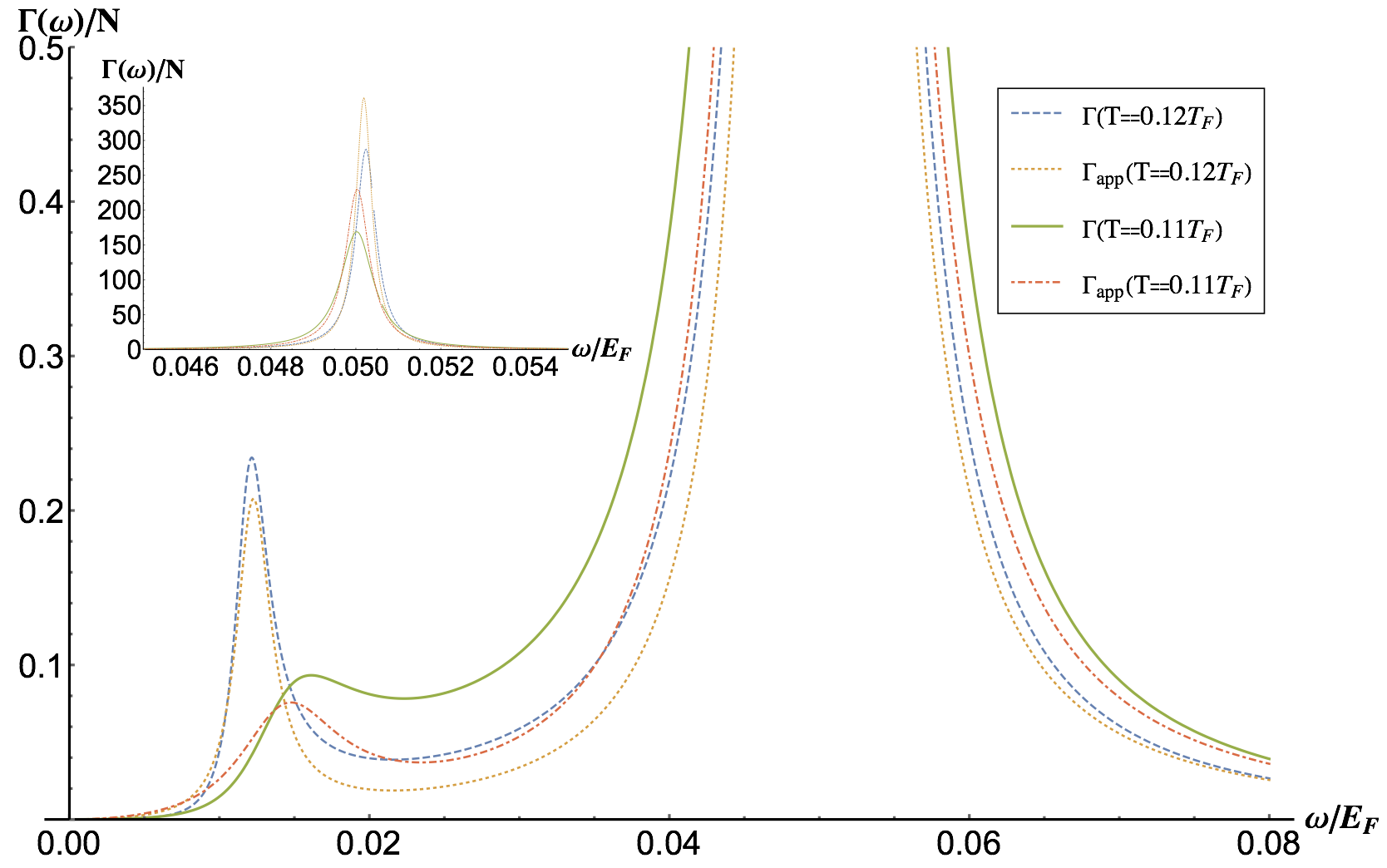}
			\caption{The energy absorption spectrum $\Gamma(\mathbf q,\omega)$ of unitary Fermi gases with $\sqrt2|\mathbf{q}|/k_F=0.1$ in the superfluid phase. The approximation $\Gamma_{\rm app}(\mathbf q,\omega)$ is shown for comparison. The inset shows the full profiles of the first sound peaks.}
			\label{Fig-low}
		\end{figure}

The kinetic coefficients can be evaluated by the variational method based on the linearized Boltzmann kinetic equation \cite{Landau kinetics,schafer kappa}. In the high temperature limit, the scattering cross section between the merely non-interacting fermions is governed by the thermal wave-length as the scattering length $a_s$ is divergent; this variational method yields
\begin{align}
\kappa=&\frac{225}{128\pi^{1/2}}m^{1/2}T^{3/2},\label{kh}\\
\eta=&\frac{15(mT)^{3/2}}{32\pi^{1/2}},\label{eh}
\end{align}
for the unitary Fermi gas \cite{schafer kappa}. In the low temperature limit, phonons are the primary excitations; we  model the system by a quantum action which includes the phonon vertex at the tree-level \cite{effective phonon}. By this model, we find
	\begin{align}
	&\kappa=\frac{5.6\times 10^{-3}}{\xi^2}\kappa_0\left(\frac{T}{T_F}\right)^2,\label{kl}\\
	&\eta=2.5\times10^{-6}\xi^{7/2}\eta_0\left(\frac{T_F}{T}\right)^5,\label{el}
	\end{align}
with $\kappa_0=m^{1/2}T_F^{3/2}$ and $\eta_0=(mT_F)^{3/2}$. The Bertsch parameter $\xi$ is experimentally found to be about $0.37$ \cite{Zwierlein2012}. In deriving Eqs.~(\ref{kl}) and (\ref{el}), we have taken the $\epsilon$-expansion result for the next-to-leading order parameters of the quantum action \cite{c1c2}. Note that Ref.~\cite{schafer eta} gives a result differing from Eq.~(\ref{el}) by a factor $2$, which results from the symmetry factor for scattering between the identical phonons. The result of \cite{schafer kappa} is about $10$ times smaller than Eq.~(\ref{kl}); we do not know at this point what gives rise to this numerical difference. The bulk viscosity $\xi_3$ is also non-zero below the superfluid transition temperature. It has been found by the same method \cite{bulk}
	\begin{align}
	&\xi_3=\frac{8.4\times 10^{-3}}{\xi^3}\xi_{30}\left(\frac{T}{T_F}\right)^3\label{xi3}
	\end{align}
with $\xi_{30}=m^{-1/2}T_F^{-3/2}$. 

\begin{figure}[t]
			\includegraphics[width=3 in]{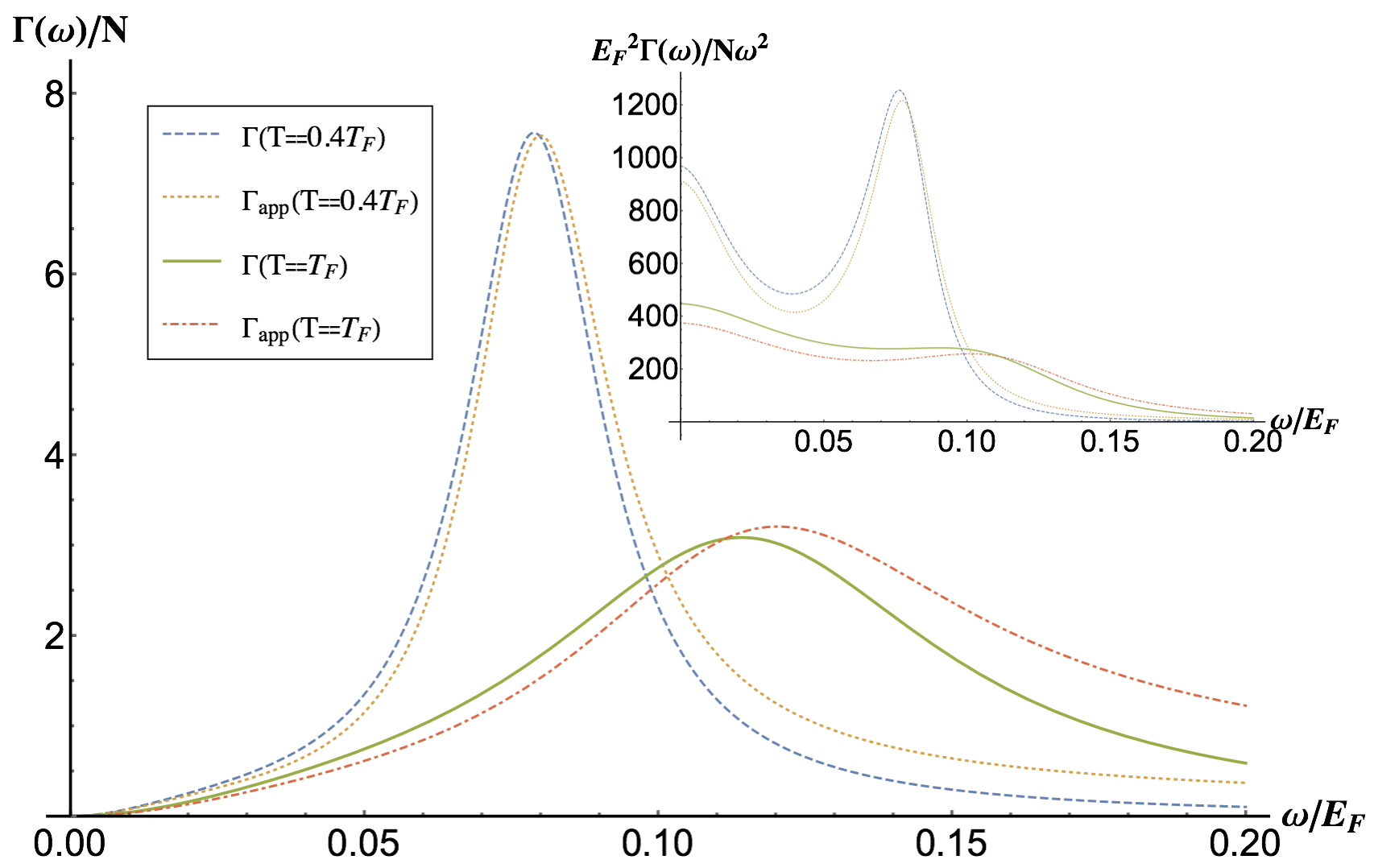}
			\caption{The energy absorption spectrum $\Gamma(\mathbf q,\omega)$ of unitary Fermi gases with $\sqrt2|\mathbf{q}|/k_F=0.1$ in the normal phase. The approximation $\Gamma_{\rm app}(\mathbf q,\omega)$ is shown for comparison. The inset plots $\Gamma(\mathbf q,\omega)/\omega^2$, which shows the Rayleigh peak at zero frequency.}
			\label{Fig-high}
		\end{figure}

By the experimentally measured equation of state \cite{Zwierlein2012} and the superfluid density \cite{Grimm2013} and assuming 
Eqs.~(\ref{kh}) to (\ref{xi3}) for the kinetic coefficients extrapolated to regions either above or below the transition temperature respectively, we plot the profiles of the the energy absorption spectrum $\Gamma(\mathbf q,\omega)$ in Figs. (\ref{Fig-low}) and (\ref{Fig-high}). We choose $\sqrt2|\mathbf q|/k_F=0.1$ for the plot. The choice falls in the regime that the expected peaks of $\Gamma(\mathbf q,\omega)$ would appear at frequencies though low but still comparable to the temperature $T$. Otherwise, if $|\mathbf q|/k_F$ is too small, the peak heights would be exponentially suppressed due to the factor $\omega e^{-\omega/T}$ in Eq.~(\ref{g-chi}). 
Figure (\ref{Fig-low}) shows that below the superfluid transition temperature $T_C$, the second peak at lower frequency, which corresponds to the second sound in the superfluid, has a much smaller magnitude compared with that of the first sound peak; this is because in the incompressible limit, i.e., $c_p/c_v\to1$, the second sound becomes purely a temperature wave which does not couple to density fluctuations \cite{Landau Hydro}. When the temperature $T$ is lowered, the superfluid fraction grows rapidly \cite{Grimm2013}, the second sound velocity increases, and so does $\eta$. As a result the second sound peak in $\Gamma(\mathbf q,\omega)$ moves to higher frequency; the two peaks come closer to each other, and at the same time become wider. When the two peaks are well separated, the plot of $\Gamma(\mathbf q,\omega)$ shows a good agreement with the approximated spectrum $\Gamma_{\rm app}(\mathbf q,\omega)$, which corresponds to the two-peak approximation result $\chi''_{\rm app}(\mathbf q,\omega)$. From Eqs.~(\ref{kl}) to (\ref{xi3}), the widths of the two peaks of the the energy absorption spectrum $\Gamma(\mathbf q,\omega)$ are both expected to be dominated by the contribution from the shear viscosity $\eta$ for $T/T_F\lesssim
 0.15$. 
Figure (\ref{Fig-high}) shows $\Gamma(\mathbf q,\omega)$ for temperatures $T$ well above the superfluid transition temperature $T_C$. In this case, $\Gamma_{\rm app}(\mathbf q,\omega)$ again gives a rather good approximation. Only the first sound peak shows up in the spectrum. In the inset, we plot $\Gamma(\mathbf q,\omega)/\omega^2 $, in which the Rayleigh peak at zero frequency is discernible; the extraction of $\kappa$ is achievable. 

In conclusion, we propose to use a simple probe, the energy absorption spectroscopy, to measure the kinetic coefficients of unitary Fermi gases in a uniform potential. We estimate that the observation of the effects of the kinetic coefficients on the energy absorption spectrum shall be within the reach of current experiments. Our method can be generalized, for example, to measure spin diffusion coefficients \cite{Zwierlein2011,Kohl2013,Thywissen2014,Thywissen2015,Thywissen2017} if perturbation potentials are chosen to be spin selective. The measurement of kinetic coefficients of uniform unitary Fermi gases would advance our understanding of strongly interacting systems and provide improved comparison with theoretical predictions.

We thank Haibin Wu and Yanhua Hou for discussions. This research was supported in part by NSFC Nos.
11474179, 11722438 and 91736103.

\emph{Note added.} After the posting of this work in the arXiv, we were notified by Martin Zwierlein at MIT that his group had already performed an experiment similar to the one proposed here, obtaining kinetic coefficients of unitary Fermi gases in uniform traps from the decay of sound modes \cite{Zwierlein_talks}. Recently Ref.~\cite{hu2017} evaluated the dynamic structure factor of unitary Fermi gases derived from the two-fluid hydrodynamic equations as well, instead together with the input of the experimentally measured $\eta$ and the high temperature form of $\kappa$. In the overlapping temperature regime, their and our results are in good agreement.

\end{document}